\documentclass[12pt]{article}
\usepackage[a4paper,top=2.5cm, bottom=2.5cm, right=2.5cm, left=2.5cm,footskip=1cm,nohead]{geometry}
\usepackage{graphicx}
\usepackage{amsmath,amsfonts,amssymb}
\usepackage{times}
\usepackage{mathptmx}
\usepackage[labelfont={bf},%
			labelsep=colon,%
			justification=raggedright]{caption}
\begin{document}
\begin{center}
	{\LARGE\textbf{Ginzburg-Landau turbulence in quasi-CW Raman fiber lasers}}
	\vskip18pt
	{\bfseries \textit{Srikanth Sugavanam$^{1}$, Nikita Tarasov$^{1,2}$, Stefan Wabnitz$^{3}$ and Dmitry V. Churkin$^{1,4,5*}$}}
	\vskip12pt
	{\small \textit{$^1$Institute of Photonic Technologies, Aston University, Birmingham, B4 7ET, UK\\
	$^2$Institute of Computational Technologies  SB RAS, Novosibirsk, 630090, Russia\\
	$^3$Dipartimento di Ingegneria dell'Informazione, Universit\`a di Brescia, and Istituto Nazionale d'Ottica del CNR, Via Branze 38, Brescia 25123, Italy\\
	$^4$Novosibirsk State University, Novosibirsk, 630090, Russia\\
	$^5$Institute of Automation and Electrometry SB RAS, Novosibirsk, 630090, Russia\\
	$^*$\underline{d.churkin@aston.ac.uk}\\
			}}
\end{center}
\vskip12pt
\begin{abstract}
	Fiber lasers operating via Raman gain or based on rare-earth doped active fibers are widely used as sources of CW radiation. However these lasers are only quasi-CW: their intensity fluctuates strongly on short time-scales. Here the framework of the complex Ginzburg-Landau equations, that are well known as an efficient model of mode-locked fiber lasers, is applied for the description of quasi-CW fiber lasers as well. The first ever vector model of a Raman fiber laser describes the experimentally observed turbulent-like intensity dynamics, as well as polarization rogue waves. Our results open debates about the common underlying physics of operation of very different laser types --- quasi-CW lasers and passively mode-locked lasers.
\end{abstract}

{\flushleft\small\textbf{keywords:} {fiber laser; rogue waves; optical turbulence; vector Ginzburg-Landau model}}

\section{Introduction}

Fiber lasers are mostly known for two well-separate regimes of operation, namely CW and short pulse mode-locked regimes. Short pulse fiber lasers generate optical solitons or dissipative pulses, and can be well modelled in terms of the complex Ginzburg-Landau equation (GLE) with a cubic nonlinear gain term and quintic nonlinear gain saturation term~\cite{grelu2012}. There is also another mode of operation of pulsed fiber lasers, namely the weak or partially mode-locking regime, where noise-like pulses are generated~\cite{horo97}. In experiments, noise-like pulses could be easily achieved by elongating the cavity, which results in a number of operational regimes depending strongly on cavity length, polarization controller settings and pump power \cite{Kobtsev3}. By modelling mode-locked lasers within the GLE framework in which cubic nonlinear gain is replaced by a nonlinear loss term, one may obtain a variety of mode-locked regimes, including those of very irregular time dynamics which could be referred to as turbulent emission~\cite{wab14}.

On the other hand, there is a wide class of CW fiber lasers based on rare-earth doped fibers or stimulated Raman scattering. In particular, Raman fiber lasers (RFLs) are designed and used as CW laser sources in many applications, including those for secure key distribution \cite{TurLPR}, and could outperform solid-states lasers in the visible range \cite{SemiLPR}: e.g., as astronomic laser guide star \cite{LPR14}. In spite of the fact that RFLs are usually designed as CW sources, their radiation is highly stochastic, exhibiting fast intensity fluctuations with extreme value statistics~\cite{Chur11,rand12}. Such quasi-CW radiation could be effectively described within the nonlinear Schr\"odinger equation (NLSE) approach \cite{Chur12}.

Here we show that CW Raman fiber lasers could be surprisingly described within the framework of the complex Ginzburg-Landau equations, which is well known as an efficient model for mode-locked lasers. In particular, we develop the first ever vector model of a RFL, that describes its turbulent-like intensity dynamics, as well as our observations of polarization rogue wave generation in the quasi-CW regime. These results open debates about the common underlying physics of operation of such different laser types --- quasi-CW lasers and passively mode-locked lasers.

\section{Quasi-CW turbulent generation}

\begin{figure}[thb]
\centering\includegraphics[width=12cm]{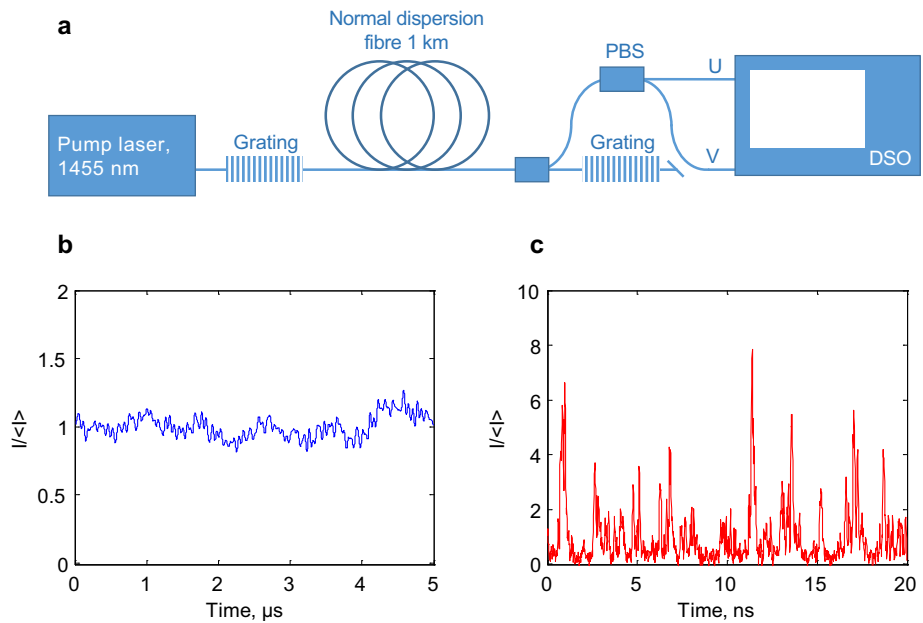}
\caption{Quasi-CW turbulent generation in a conventional Raman fiber laser measured with different oscilloscope's bandwidth and at different time scales.
}
\label{Fig1}
\end{figure}

In our experiments, we use a linear cavity RFL made of 1 km long normal dispersion fiber and two fiber Bragg gratings (FBGs). The laser is pumped at 1455 nm and generates at 1550 nm. The time dynamics is analyzed by a 33 GHz bandwidth real-time oscilloscope. When observed on $\mu s$ time scale (oscilloscope bandwidth is reduced to 1 GHz), the intensity dynamics exhibits relatively small fluctuations, and the laser may be considered a CW source, Fig.\ref{Fig1}a. On $ns$ time scale, the laser exhibits highly stochastic dynamics, Fig.\ref{Fig1}b. Such generation regime could be described in terms of the weak wave turbulence approach \cite{Babin07}, hence it is referred to as \textit{turbulent generation}. Moreover, in analogy with flows in pipes, laminar generation (a regime of a large number of correlated modes with suppressed total intensity fluctuations) was also observed  \cite{chur13}.

We further experimentally characterize the radiation in spatio-temporal domain \cite{chur13}; similar techniques are also used in studies of various pulsed fiber systems \cite{Coen,Broderick15,Barland}. We measure in real-time how the intensity pattern, $I(t)$, evolves slowly over a number $N$ of consecutive cavity round-trips.  We found that the radiation has a distinct spatio-temporal structure, $I(t,N)$, Fig.\ref{Fig2}a, with regions of high intensity being quasi-stationary of up to 20-30 cavity round-trips, and sitting on a background of almost zero intensity. The typical nonlinear length in our case is only around one tenth of the cavity round-trip. No indication of mode-locking is observed: neither in temporal dynamics, nor in spatio-temporal intensity dynamics, or the in radio-frequency spectrum.

It is important to note that passive mode-locking could be achieved in RFLs by using nonlinear polarization evolution and special cavity design \cite{RamanNPE}. In our case, the laser has the simplest linear cavity design, and does not comprise any special elements that could possibly lead to the arising of mode-locking: all components are made from non-PM fiber, and there are neither polarization beam splitters nor polarization controllers. The pump light is unpolarized, and we have experimentally checked that the polarization of the generated Stokes wave is random.

\section{Description of quasi-CW radiation within vector Ginzburg-Landau model}

In spite of the fact that the laser operates in the quasi-CW regime, in the following we use the Ginzburg-Landau equation model to describe its light emission.
The GLE is derived by averaging the propagation of a Stokes field over each round trip in the cavity \cite{HTW1}:
\begin{eqnarray}
t_r\partial_{t'} E_s & = & \frac{\left(G_pL-T^2\right)}{2}E_s+L\left(\beta_0+i\frac{\beta_2}{2}\right)\partial_{\tau'}^2E_s  \nonumber\\
 & - & L\left(\frac{\sigma_0}{2}-i\gamma_0\right)|E_s|^2E_s
\label{GLEdim}
\end{eqnarray}
\noindent where $t'$ is a continuous (slow) temporal variable that replaces the round-trip number $N$, $\tau'$ is a retarded time equivalent to time coordinate $t$ measured in experiments, $E_s$ is the complex field envelope of Stokes light, 
and $t_r$ is the round-trip time for a cavity of length $L$. Here $T$ is the output coupler amplitude transmission coefficient. Raman gain is introduced by a factor $G_p=g_p\left\langle |E_p|^2\right\rangle$, where $g_p$ is the Raman gain coefficient and $E_p$ is the pump field. 
\begin{figure}[thb]
\centering\includegraphics[width=12cm]{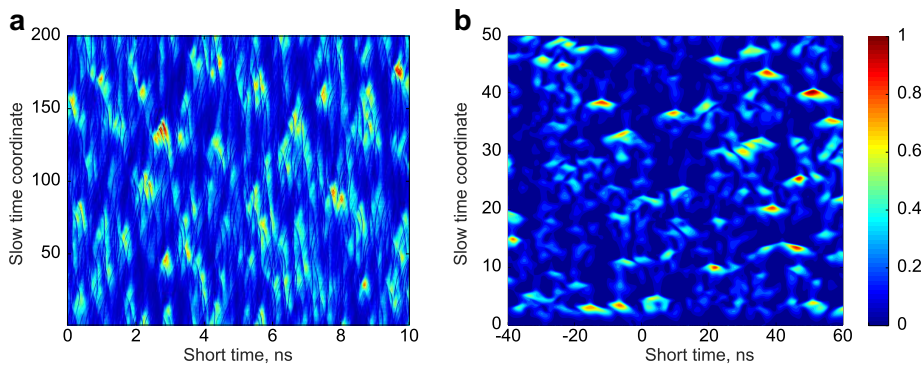}
\caption{Spatio-temporal turbulent intensity dynamics, $I(t,N)$ of quasi-CW Raman fiber laser in (a) experiment and (b) numerical simulations within vector Ginzburg-Landau model.} 
\label{Fig2}
\end{figure}

Brackets denote spatial average over a length scale $L_a$, that is much larger than the dispersion walk-off length $L_w$ between pump and Stokes waves, and much shorter than the nonlinear length~\cite{Chur12}. Bandwidth limited gain originates from the finite spectral bandwidth of the FBG-based mirrors which is smaller than the width of Raman gain spectral profile.  Mirrors profiles are represented by the parabolic approximation $G(\omega)=G_p\left(1-(4/B^2)(\omega-\omega_0)^2\right)$, so that $\beta_0=2G_p/B^2$, where $\omega_0$ is the central frequency of the FBGs spectral profiles. Cavity average group velocity dispersion and Kerr nonlinearity are taken into account via coefficients $\beta_2$ and $\gamma_0$. Note that the GLE requires the phenomenological introduction of a nonlinear saturation of the gain, which is described here by the coefficient $\sigma_0>0$.

It is convenient to recast Eq.(\ref{GLEdim}) in dimensionless units:
\begin{equation}
 \partial_tU =  U+\left(1+i\beta\right)\partial_\tau^2U -\left(1-i\gamma\right)|U|^2U
\label{GLE}
\end{equation}
where $t=\hat{g}t'/2t_r$,  $\hat{g}=g_0L-T^2$, $\tau=\tau'\sqrt{\hat{g}/2\beta_0L}$,  $U=\sqrt{\sigma_0L/\hat{g}}E$, $\beta=-\beta_2/2\beta_0$, and $\gamma=2\gamma_0/\sigma_0$.

Since the laser does not contain any polarization-selective elements, one needs to consider a vector extension of the Eq.~(\ref{GLE}). Besides nonlinear gain saturation and spectral filtering, in the vector case it is necessary to include cross-phase modulation, as well as cross-gain saturation between the two orthogonally polarized laser modes. We may thus describe the dissipative dynamics of a vector CW laser in terms of the coupled GLEs~\cite{Lec14}:
\begin{eqnarray}
 \partial_tU & = & U+\left(1+i\beta\right)\partial_\tau^2U-\left(1-i\gamma\right)|U|^2U-\left(\eta-i\rho\right)|V|^2U
 \nonumber \\
 \partial_tV & = & V+\left(1+i\beta\right)\partial_\tau^2V-\left(1-i\gamma\right)|V|^2V-\left(\eta-i\rho\right)|U|^2V \nonumber 
\end{eqnarray}
where now $U(t,\tau)$ and $V(t,\tau)$ represent the two orthogonal polarization components of the field in the cavity. Here $\eta$ and $\rho$ describe the action of nonlinear polarization cross gain-saturation and phase modulation, respectively.

We set $\beta=-0.5$ to consider the normal dispersion regime. It is necessary that $\gamma,\rho \gg 1$ to be in the limit where the coupled GLEs approach could be reduced to perturbed vector NLSE. For example, we take $\gamma=\rho=25$. In addition, we set the cross-gain saturation $\eta=0.4$.
In other words, we operate in a regime where the conservative nonlinear length associated with the Kerr nonlinearity, $L_{nl}^C\propto \gamma^{-1}$, is much shorter than the dissipative length associated with linear gain and nonlinear gain saturation, $L_{nl}^D=1$. As initial condition, we used a weak random noise seed with normal distribution in each polarization component.

We found a good qualitative agreement between the experimentally measured spatio-temporal dynamics of the quasi-CW fiber laser and the numerically calculated dynamics within the developed vector phenomenological model based on the coupled GLEs, Fig.~\ref{Fig2}b. This clearly shows that the \textit{quasi-CW} fiber laser can be effectively described by a phenomenological model that is widely used for the modelling of \textit{mode-locked} lasers.

One more important result is that the developed model is the first ever full vector model of the generation in Raman fiber lasers which takes into account polarization evolution. Previously, polarization effects have been studied only within the simplest power balance model~\cite{suret2004influence}, which, obviously, does not take into account any of the effects that are induced by dispersion/nonlinearity.

\begin{figure}[thb]
\centering\includegraphics[width=12cm]{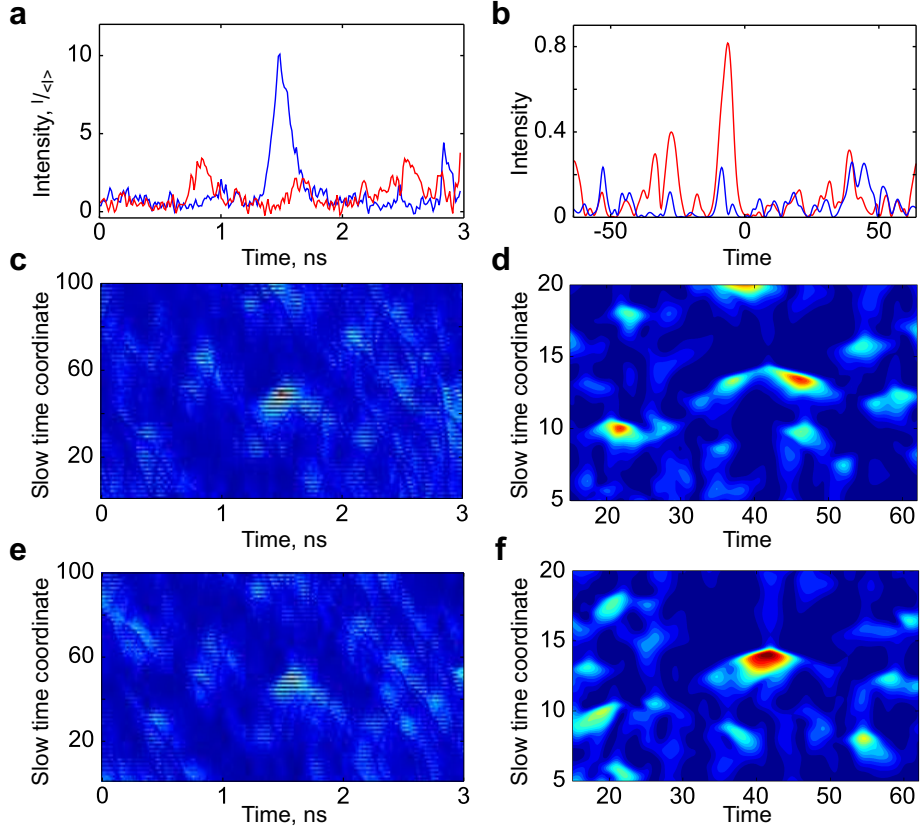}
\caption{Polarization rogue waves in CW Raman fiber laser in experiments (left column) and simulations (right column). (a,b) Intensity dynamics in orthogonal polarizations $U$ and $V$ (red and blue). (c-f) Spatio-temporal intensity dynamics in orthogonal polarizations: $U$ is shown on (c,d) and $V$ on (e,f).}
\label{Fig3}
\end{figure}

The quasi-CW RFL generates high intensity events, Fig. 1b. Those events could be associated with optical rogue waves, see recent review \cite{dudley2014instabilities}. The developed vector model provides the possibility to analyze the intensity dynamics in two orthogonally polarized modes and thus uncover the polarization nature of observed extreme events. Both experiments and simulations reveal that most of extreme events occur in a single polarization, Fig.\ref{Fig3}a,b. Namely, whenever there is an extreme event in one polarization, the intensity in the other polarization remains relatively low. Therefore there is no clear correlation or antiphase behavior that is emerging from the laser when extreme peaks are generated, in contrast with vector lasers based on rare-earth fibers~\cite{Lec14}. Note that polarization rogue waves were previously theoretically predicted within the coupled NLSE approach~\cite{Stefano12,Stefano14}.

Furthermore, we analyzed the spatio-temporal intensity dynamics in both polarizations, Fig.\ref{Fig3}c-f.
There is again a good qualitative agreement between experiments and theory. Note however that experiments exhibit horizontal ``stripes'' , which indicate the presence of a bifurcation into the period-doubling or tripling regime, i.e., the intensity patterns repeat every second or third round-trip in the cavity. These effects are beyond the limits of applicability of the coupled GLEs, which are derived in the mean-field approximation. A proper analysis of the period doubling regime may however be carried out by deriving a set of coupled GLEs for each polarization mode, in analogy with the treatment of period-doubling in synchronously pumped passive fiber cavities~\cite{haelt92}.

\section{Discussion}

Our results imply that the very essence of both quasi-CW fiber lasers and partially mode-locked fiber lasers should be revisited, as the evolution towards irregular and fast self-pulsing regimes for both RFLs and passively mode-locked lasers may be described in terms of the same universal phenomenological model --- GLEs. Indeed, in both laser types the radiation is noise-like: in quasi-CW lasers, high intensity pulses suddenly emerge from optical turbulence as isolated flashes of light, and subsequently disappear without a trace~\cite{hamma10}.

In partially mode-locked lasers, a stable pulse train can be generated because of the presence of nonlinear polarization evolution, saturable absorber mirrors or other mode-locking mechanisms. Despite that, if the laser is long enough, and depending on cavity parameters, each pulse could exhibit an irregular noise-like structure both in temporal and spectral domains \cite{Kobtsev3,rung14}. As the resulting pulse width could be orders of magnitude longer than the typical noise-like filling, these pulses could be effectively treated as quasi-CW radiation. Thereby, the formation and properties of their internal structure could be governed by mechanisms similar to those of quasi-CW lasers. The Ginzburg-Landau equation model thus provides an efficient tool for the unified and universal description of both quasi-CW and partially mode-locked fiber lasers.

There could be another option for the joint description of lasers of different types. Indeed, optical turbulence in long RFLs has been previously described in the frequency domain by means of the wave kinetic equation, thus highlighting the role of nonlinear four-wave mixing (among the large number of cavity modes) in the formation of noise-like generation~\cite{Babin07}. The proposed here GLE approach provides a dual, time-domain based description of the turbulent dynamics in quasi-CW lasers~\cite{Manne}.
In passively mode-locked lasers, which are dynamically described in time domain by GLE approach, the frequency-domain description in terms of the weak wave turbulence approach could be also fruitful. Indeed, besides the small section of the laser where the gain and mode synchronization mechanisms are important, the rest of the cavity length in long passively mode-locked fiber lasers could be potentially treated as an effectively conservative Hamiltonian system. Hence the stochastic processes leading to the noise-like filling of mode-locked pulses could be also described by means of the weak wave turbulence theory, i.e., in the frequency-domain. The origin of noise-like pulses in passively mode-locked lasers is still unknown.

Another interesting question is if the recently emerged quasi-CW random fiber lasers which are phenomenologically described within the NLSE approach \cite{SmirnovRandom} could also  be described within the vector Ginzburg-Landau model. Such description could potentially show the route to the mode-locking of random fiber lasers.

\section{Conclusion}

 We have demonstrated that the turbulent spatio-temporal dynamics of quasi-CW fiber lasers could be described within a simple, and yet universal vector model based on coupled Ginzburg-Landau equations, that so far had only been used for the description of mode-locked lasers. The first ever vector model of Raman fiber laser describes well our experimental observations of polarization rogue waves. We believe that our results will bring  insight into the joint physics behind the noise-like operation of both quasi-CW and partially mode-locked fiber lasers, and could open the way to the unified description of both laser types and dynamical instabilities in theirs generation.

\section*{Acknowledgement}

The authors acknowledge support from the ERC UltraLaser project, the Italian Ministry of University and Research (MIUR) (grant contract 2012BFNWZ2). N.T. is supported by the Russian Science Foundation (Grant No. 14-21-00110).

\end{document}